\documentclass[english]{article}
\usepackage[T1]{fontenc}
\usepackage[utf8]{inputenc}
\usepackage{amsthm}
\usepackage{amsmath}

\makeatletter
\numberwithin{equation}{section}
\numberwithin{figure}{section}
\theoremstyle{plain}
\newtheorem{thm}{Theorem}
  \theoremstyle{definition}
  \newtheorem{defn}[thm]{Definition}

\makeatother

\usepackage{babel}

\begin{document}

\title{Symmetry and Uncountability of Computation}

\date{2010-09-22}

\author{KOBAYASHI Koji}
\maketitle
\begin{abstract}
This paper talk about the complexity of computation by Turing Machine.
I take attention to the relation of symmetry and order structure of
the data, and I think about the limitation of computation time. First,
I make general problem named {}``testing problem''. And I get some
condition of the NP complete by using testing problem. Second, I make
two problem {}``orderly problem'' and {}``chaotic problem''. Orderly
problem have some order structure. And DTM can limit some possible
symbol effectly by using symmetry of each symbol. But chaotic problem
must treat some symbol as a set of symbol, so DTM cannot limit some
possible symbol. Orderly problem is P complete, and chaotic problem
is NP complete. Finally, I clear the computation time of orderly problem
and chaotic problem. And P != NP.
\end{abstract}

\section{Introduction}

This paper talk about the complexity of computation by Turing Machine.
I take attention to the relation of symmetry and order structure of
the data, and I think about the limitation of computation time.

First, I make general problem named {}``testing problem''. And I
get some condition of the NP complete by using testing problem.

Second, I make two problem {}``orderly problem'' and {}``chaotic
problem''. Orderly problem have some order structure. And DTM can
limit some possible symbol effectly by using symmetry of each symbol.
But chaotic problem must treat some symbol as a set of symbol, so
DTM cannot limit some possible symbol. Orderly problem is P complete,
and chaotic problem is NP complete.

Finally, I clear the computation time of orderly problem and chaotic
problem. And $P\neq NP$.

\section{Make {}``Testing problem''}

I make the genelic problem that have DTM and NTM problem in it.

{}``Verifier'' is the TM that can decide a problem. Verifier can
go to halting (accepting or rejecting) configuration finally. {}``Verify
data'' is the data that verifier compute. {}``Verifying problem''
is the problem that verifier decide. {}``Specific symbol'' is the
symbol that make up verify data. {}``Verify problem'' is the problem
that verifier can compute.

{}``Checker'' is the TM that can use {}``Generic symbol''. {}``Generic
symbol'' is the special symbol that mean one of the specific symbol,
and one of the generic symbol that mean fewer specific symbol. {}``Check
data'' is the data that have generic symbol. Checker reject the check
data that verifier reject all verify data that change check data's
generic symbol to any specific symbol, and checker accept other check
data. That is to say, checker accept the check data that verifier
accept one or more verify data that change check data's generic symbol
to any specific symbol. {}``Checking problem'' is the problem that
checker can decide. 

{}``Selected symbol'' is the specific symbol (or the generic symbol
that mean fewer specific symbol) that can change from the generic
symbol in check data.

Checker is the Verifier that have transition function to use specific
symbol. If check data don't have specific data, checker compute like
verifier. And checker decide same (accepting or rejecting) configuration.

{}``Testing problem'' is verifying problem and checking problem.
{}``Tester'' is verifier and checker. {}``Test data'' is verify
data and check data.

{}``Computation result'' is halting (accepting or rejecting) situation
that tester compute test data. And {}``remaining configuration''
is the configuration to clear the situation that tester do not decide
accept or reject.

It's some relation between testing problem and verifying problem.
\begin{thm}
\label{thm:Testing problem and verifying problem}If verifying problem
is P, testing problem is NP-complete.\end{thm}
\begin{proof}
It's easy to proof that testing problem is in NP.

If a test data is acceptable, some verify data is acceptable that
match any specific symbol with the testing data. The verifying problem
is P that verify the verifying data, so testing problem is in NP.

To proof testing problem is NP-hard, I clear that SAT can reduce to
testing problem with polynomial time.

Those test data have SAT's logical formula as specific symbol, and
SAT's truth value as general symbol. The test data have polynomial
length of logacal fomula because truth value is shorter than logical
fomula. And It's can polynomial time to get variable from logacal
fomula. So SAT can reduce to testing problem with polynomial time.

So, if verifying problem is P, testing problem is NP-complete.
\end{proof}

\section{Symmetry and order of the problem}

To talk about complexity of problem, I make two testing problem.

\subsection{Orderly problem}

First, I make the problem that can separate genelic symbol each other.
\begin{defn}
\label{def:Orderly problem} ''Orderly problem'' is the testing
problem that all test data have at least one of the selected symbol
fix the computation result, and tester can get selected symbol with
polynomial time. {}``Focus symbol'' is the generic symbol, {}``Symmetry
symbol'' is the selected symbol that the computation result fix.
And {}``Simplify symbol'' is the selected symbol that keep same
computation result as focus symbol.
\end{defn}
Orderly problem's test data have at least one of selected symbol that's
coputation result is symmetry. Simplify symbol have same computation
result that other generic symbol is any selected symbol. And to use
this symmetry, tester can change from generic symbol to selected symbol
orderly. If the computation reslut is reject, tester change the generic
symbol to other generic symbol that do not have the selected symbol
in it with keeping computation result. If the computation reslut is
accept, tester change the generic symbol to selected symbol with keeping
computation result. That is to say, Focus symbol have free part from
other generic symbol at point of view of the computation result.

If Tester can decide accept/reject/remind when focus symbol change
any selected symbol, tester can decide symplify symbol in polynomial
time. Because selected symbol count is limited, tester can decide
accept/reject/remind of all selected symbol in polynomial time.

And, test data that change from focus symbol to simplify symbol is
keep condition of orderly problem. So I can say that;
\begin{thm}
\label{thm:Order of the generic symbol of orderly problem}Generic
symbol in test data of orderly prblem have at least one of the order
structure.\end{thm}
\begin{proof}
Because of \ref{def:Orderly problem}, test data of orderly problem
have at least one of the focus symbol, and tester can change from
focus symbol to simplify symbol. And the test data that change simplify
symbol also have focus symbol and simplify symbol. And this can do
till all generic symbol change to specific symbol. So, all generic
symbol have some order to change specific symbol.\end{proof}
\begin{thm}
\label{thm:Complexity of orderly problem}If verifying problem is
P, orderly problem is P-complete.\end{thm}
\begin{proof}
First, I clear orderly problem is in P. Because of \ref{thm:Order of the generic symbol of orderly problem},
tester of orderly problem can change from every focus symbol to simplify
symbol in polynomial time. And tester can change from test data to
verify data in polynomial time (because test data have shorter length
generic symbol than the test data length). And tester can decide computation
result of verify data in polynomial time, so tester can compute all
in polynomial time. 

To proof orderly problem is P-hard, I clear that HornSAT can reduce
to orderly problem with log space.

Like \ref{thm:Testing problem and verifying problem}, tester can
make from HornSAT to test data. Tester need only logical fomula's
current position to get variable. So HornSAT can reduce to testing
problem with log space.

I clear that Test data of HornSAT have orderly problem's condition.
HornSAT's variable have some order. Tester can put variable value
that is necessary to accept test data. Finally tester can decide computation
result. This is same condition HornSAT and Test data of HornSAT, tester
can change from generic symbol as variable to specific symbol as value.
So test data of HornSAT is orderly problem's test data.

So, if verifying problem is P, orderly problem is P-complete.
\end{proof}

\subsection{Chaotic problem}

Second, I make the problem that is necessary to compute genelic symbol
as a large body.
\begin{defn}
\label{def:Chaotic problem}''Chaotic problem'' is the testing problem
that tester cannot dicede computation result by changing any one generic
symbol to selected symbol. If tester decide computation result, tester
must decide many generic symbol to selected symbol. And such generic
symbol is not limited. {}``Extended symbol'' is the generic symbol
above. {}``Unit symbol'' is each generic symbol that make extended
symbol. {}``Extended symbol length'' is count of extended symbol.
Extended symbol that change from partly or all generic symbol to selected
symbol is Extended symbol. Extended symbol is out of count. 

{}``Universal problem'' is the co-problem of chaotic problem.
\end{defn}
Chaotic problem do not have good condition to decide focus symbol
and simplify symbol, so tester must decide another generic symbol
to decide any generic symbol. Tester must compute some generic symbol
as a large mass, cannot compute like orderly problem. Tester must
compute extended symbol as a incleasing generic symbol or specific
symbol, so tester must compute incleasing symbol that come exponent
size of extended symbol length. And each unit symbol that make extended
symbol is symmetry for tester. If tester break the symmetry by using
some unit symbol first, chaotic problem keep symmetry when all test
data get together.

By the way, if extended symbol length is limited, tester can compute
the problem as orderly problem because tester can compute the extended
symbol change to new generic symbol or specific symbol.

Chaotic problem is the set of testing problem cleaning the orderly
problem. If orderly problem is testing problem, chaotic problem is
empty. If testing problem have the problem that is not in orderly
problem, the problem is chaotic problem.
\begin{thm}
\label{thm:Being chaotic problem} If verifying problem is P, chaotic
problem is not empty.\end{thm}
\begin{proof}
To proof this theorem, I use reduction to absurdity. I suppose that
chaotic problem is empty if verifying problem is P. So, all test data
have focus symbol and symmetry symbol, and the testing data that change
from focus symbol to symmetry symbol have same computation result.

But testing problem is NP-complete having SAT, so test data do not
have computation result symmetry each test data like orderly problem.
For example, SAT have any logical fomula that make given truth table.
So, SAT have any logical fomula that result is true and false if a
part of variable fix the value. Testing data from SAT do not have
focus symbol and symmetry symbol, and testing data is not orderly
problem.

That is to say, This assumption conflict the testing problem condition.

So, if verifying problem is P, chaotic problem is not empty.
\end{proof}

\section{Difference between Orderly problem and Chaotic problem}

I clear the difference of complexity of chaotic problem and orderly
problem.
\begin{thm}
\label{thm:Chaotic problem's complexity}If verifying problem is P,
chaotic problem is not in P.\end{thm}
\begin{proof}
To proof this theorem, I use reduction to absurdity. I suppose that
chaotic problem is in P if verifying problem is P.

Some testing data A and set of testing data that change testing data
A of a generic symbol to a selected symbol. Tester can get All testing
data's computation result in polynomial time. And tester can get All
testing data that have same computation result with testing data A.
So, tester can get testing data with same computation result testing
data A and change from a generic symbol to seletcetd symbol. Testing
data A have condition of focus symbol and simplify symbol. And testing
data A is any testing data of chaotic problems.

But, if testing data A is chaotic problem's data, this generic symbol
and selected symbol is not independ other generic symbol. So tester
can only change from generic symbol to selected symbol with other
generic symbol, and can not change only one generic symbol without
changing other generic symbol. Because unit symbol that make Extended
symbol is symmetry to tester, if tester use some unit symbol first,
chaotic problem keep symmetry when all test data get together. So
tester must use all extended symbol as free symbol each other. Extended
symbol count is exponent size of extended symbol length, so tester
can not use extended symbol as generic symbol and specific symbol
same polynomial time.

If this generic symbol can change selected symbol at any extended
symbol, the generic symbol is free from extended symbol. And if all
test data have such generic symbol, testing problem do not have extended
data. So This assumption conflict the chaotic problem condition.

So, if verifying problem is P, chaotic problem is not in P.
\end{proof}
If the verifying problem is P, the orderly problem is P-complete and
the chaotic problem is NP-complete. So, I clear $P\neq NP$.
\begin{thm}
$P\neq NP$
\end{thm}

\section{More detail about difference between orderly problem and chaotic
problem}

I talk about chaotic problem's computation time in more detail.

To clear the limit of the orderly problem, I make special orderly
problem.
\begin{defn}
\label{def:Saturated orderly problem}''Saturated orderly problem''
is the orderly problem that do not have all the data that make some
generic symbol and some specific symbol, and the count of data is
not limited.\end{defn}
\begin{thm}
\label{thm:Being saturated orderly problem}Saturated orderly problem
is not empty.\end{thm}
\begin{proof}
To proof this theorem, I use the testing problem from CNFSAT.

Same as \ref{thm:Testing problem and verifying problem}, I make testing
problem from CNFSAT. At this testing problem, logical fomula is specific
symbol and true value is generic symbol. 

Now, to change generic symbol to specific symbol, I change CNFSAT
to HornSAT. I can change CNFSAT to HornSAT by putting false to some
variable. The variable is over positive literal of each clause. And
I can do same operation to testing problem from CNFSAT. The testing
problem that do same operation is like HornSAT and fill orderly problem
condition.

But, the testing problem like HornSAT have the variables that force
false, so testing problem do not have all data of generic symbol and
specific symbol. And the count of structure is not limited.

So, The orderly problem like this is saturated orderly problem.
\end{proof}
Saturated orderly problem is the limit of orderly problem. To clear
difference between chaotic problem and orderly problem, I make new
problem.
\begin{defn}
\label{def:Saturated chaotic problem}''Saturated chaotic problem''
is the testing problem that's verifying problem is saturated orderly
prblem.\end{defn}
\begin{thm}
\label{thm:Being saturated chaotic problem}Saturated chaotic problem
is not empty.\end{thm}
\begin{proof}
The testing problem that's verifying problem is saturated orderly
prblem. The saturated orderly problem's generic symbol is the testing
problem's specific symbol. And It is same computation result that
change the saturated orderly problem's generic symbol to testing problem's
generic symbol.

But, saturated orderly problem do not have all data of generic symbol
and specific symbol because of \ref{def:Saturated orderly problem}.
So, if I group all test data that match a part of data, some group
have only the test data that is different specific symbol part only.
Saturated orderly problem do not have test data that change the part
of the specific symbol to generic symbol.

So, the testing problem that's verifying problem is saturated orderly
problem have the data that is not in the saturated orderly problem.
And testing problem is not saturated orderly problem.
\end{proof}
Saturated chaotic problem is complexer than saturated orderly problem.
Tester compute saturated chaotic problem in exponent time.
\begin{thm}
\label{thm:Complexity of saturated chaotic problem}If Saturated orderly
problem's verifying problem is P, tester compute saturated chaotic
problem in exponent time of extended symbol size.\end{thm}
\begin{proof}
I make the test data that have the generic symbol. Saturated orderly
problem do not have the generic symbol, and saturated chaotic problem
have the generic symbol. So, it is not orderly problem that can computate
the generic symbol without changing to specific symbol. And the verifying
problem that change the generic symbol to specific symbol is P-complete
because the saturated orderly problem's verifying problem is P. Count
of the extended symbol that maked the generic symbol is amount exponent
size, so the computation time of the saturated chaotic problem is
exponent size of extended symbol size.
\end{proof}
Now, if saturated orderly problem's verifying problem is P, saturated
orderly problem is P-complete and saturated chaotic problem is NP-complete.
And because of \ref{def:Saturated orderly problem} and \ref{thm:Complexity of saturated chaotic problem},
saturated chaotic problem's extended symbol size is not limited, and
computation time is exponent size of extended symbol size. So, I clear
$P\neq NP$.
\begin{thm}
$P\neq NP$
\end{thm}

\end{document}